\documentclass[12pt,draftcls,onecolumn]{IEEEtran}
\IEEEoverridecommandlockouts
\usepackage{cuted}
\usepackage{cite}
\usepackage{bm}
\usepackage{amsmath,amssymb,amsfonts}
\usepackage{algorithm}
\usepackage{algorithmic}
\usepackage{graphicx}

\usepackage{graphicx}
\usepackage{subcaption}
\usepackage{makecell}
\usepackage{diagbox}

\usepackage[letterpaper, left=0.625in, right=0.625in, bottom=1.02in, top=0.70in]{geometry}

\usepackage{textcomp}
\usepackage{xcolor}

\newcommand{\be}{\begin{equation}}
\newcommand{\ee}{\end{equation}}

\def\BibTeX{{\rm B\kern-.05em{\sc i\kern-.025em b}\kern-.08em
    T\kern-.1667em\lower.7ex\hbox{E}\kern-.125emX}}

\begin{document}

\title{Physics-informed Generalizable Wireless Channel Modeling with Segmentation and Deep Learning: Fundamentals, Methodologies, and  Challenges
}
\author{\IEEEauthorblockN{Ethan Zhu\IEEEauthorrefmark{1}, Haijian Sun\IEEEauthorrefmark{2}}, Mingyue Ji\IEEEauthorrefmark{3} \\
\IEEEauthorblockA{
\IEEEauthorrefmark{1}Green Canyon High School, 2960 Wolf Pack Wy North, North Logan, UT USA. \\
\IEEEauthorrefmark{2}School of Electrical and Computer Engineering, University of Georgia, Athens, GA  USA. \\
\IEEEauthorrefmark{2}Department of Electrical and Computer Engineering, University of Utah, Salt Lake City, UT  USA. \\
Emails: \IEEEauthorrefmark{1}ethanzhu987@gmail.com, \IEEEauthorrefmark{2}hsun@uga.edu,
\IEEEauthorrefmark{3}mingyue.ji@utah.edu
}}
\maketitle

\begin{abstract}
Channel modeling is fundamental in advancing wireless systems and has thus attracted considerable research focus. Recent trends have seen a growing reliance on data-driven techniques to facilitate the modeling process and yield accurate channel predictions. In this work, we first provide a concise overview of data-driven channel modeling methods, highlighting their limitations. Subsequently, we introduce the concept and advantages of physics-informed neural network (PINN)-based modeling and a summary of recent contributions in this area. Our findings demonstrate that PINN-based approaches in channel modeling exhibit promising attributes such as generalizability, interpretability, and robustness. We offer a comprehensive architecture for PINN methodology, designed to inform and inspire future model development. A case-study of our recent work on precise indoor channel prediction with semantic segmentation and deep learning is presented. 
The study concludes by addressing the challenges faced and suggesting potential research directions in this field.
\end{abstract}


\newpage

\section{Introduction}
Wireless channel modeling and estimation are instrumental for many applications in wireless communication, such as localization, frequency selection, coordinated site deployment, and antenna design. It is not surprising that channel modeling has attracted extensive research interests since the advent of wireless systems.  A good understanding of how radio propagates through space and precise estimation of their parameters provide a foundation for wireless communication system optimization and performance evaluation. For example, precise knowledge of signal multipath component (MPC) between the transmitter (TX) and receiver (RX) allows for the implementation of advanced transmission strategies like pre-Rake at the TX side, which helps to mitigate multi-path effects. Similarly, this knowledge can aid in implementing more efficient and less complex decoding techniques at the RX side. 


The mathematical representation of a wireless channel, viewed as a linear time-varying system, is characterized by its channel impulse response (CIR) in the time domain that consists of a total of $K$ MPCs  as $h(t, \mathbf{\Theta}, \mathbf{\Phi}) = \sum_k a_k(t) \delta (t - \tau_k(t))   \delta (\mathbf{\Theta} - \mathbf{\Theta}_k(t))  \delta (\mathbf{\Phi}  - \mathbf{\Phi}_k(t))$, where $a_k(t)$ is the complex attenuation, $\tau_k(t)$ is the delay (time-of-flight), $\mathbf{\Theta}_k(t)$ and $\mathbf{\Phi}_k(t)$ are angle of departure and arrival (AoD, AoA) of $k$-th path, respectively. Formally,  the goal of wireless channel or radio propagation modeling, is to estimate and predict those parameters (either instantaneous or statistics) in each path, given the TX and RX location and the environment they reside in. 
Two primary methods exist for channel modeling, which are succinctly introduced here.
\subsection{Stochastic channel models}
In early days, knowledge of radio propagation theory usually starts with extensive field measurements in a specific scenario. Through regression analysis, the collected data are used to identify trends and patterns that can be fit and incorporated into mathematical models. 
Many classical models are derived from above steps, such as Okumura and Walfisch-Bertoni model for urban,  Hata-Davidson and Walfisch-Ikegami model for suburban. 
Due to measurement device and processing limitations, CIR cannot be directly obtained. Instead, received signal strength (RSS) is often used to characterize channel quality. While CIR provides a comprehensive features of wireless channel, RSS is a simple measure of its aggregated effects. They are intrinsically related and can be transformed through bandwidth, signal waveform, and carrier frequency.  
Stochastic models are parameterized and relatively straightforward, they are limited in their ability to account for complex geometries and dynamic environmental factors, rendering them less suitable for applications requiring high accuracy. 

\subsection{Deterministic channel models}
Electromagnetic waves follow basic physics laws that can reflect, scatter, or diffuse when encountering an object. Therefore, theoretical models such as Maxwell's equations can describe those behaviors. Many simulators are built to characterize a more deterministic behavior of wireless signals within near-field, in the broad area of computational electromagnetics (CEM). Methods include the finite-difference time-domain method, or the vector parabolic equation method.  
For far-field scenarios such as wireless system, ray tracing (RT) is usually used. By launching a large number of rays following TX antenna pattern, RT can trace individual ray with physical behaviors from CEM when interacting with obstacles. The RX side will collect all reachable rays for MPC parameters calculation.   
Although  equations in CEM can be reformulated as partial differential equations, where numerical solutions can simplify the computation process, it is still resource consuming. The same applies to RT, especially in complex environment. Besides, deterministic approach requires detailed geometry of the propagation environment, including the 3D dimension, covering materials of each object, even smoothness of the surface, etc.,  which further hinders its practicability. Lastly, RT is considered as ``one-shot'', even slight changes in the environment requires a complete re-run of the simulation to accurately reflect the new conditions. 

To date, it is still challenging to strike the balance between stochastic and computational models. Research in this field has predominantly focused on finding a trade-off approach that can provide good accuracy and real-time capability, yet requires less efforts in geometry construction and computations. Topics include the hardware acceleration of RT interactions with mesh materials, the development of approximations for geometry-specific models such as cluster-based or probabilistic models, and the simplification of complex mathematical equations.


Machine learning (ML) is emerged as a promising technique in several major socioeconomic sectors and has found wide applications and success in computer vision (CV), autonomous driving, and recently the large language model (LLM). The breakthrough of ML has also fueled increasing interests in their applications on channel modeling. ML shows its powerful features in extracting implicit mapping between input and output data, which is particularly fit for signal propagation under complex scenarios, where their mapping is difficult to be established through equations. Existing works have applied ML for not only channel characterization such as identification of LoS or NLoS, but also model prediction when given a new propagation scenario \cite{huang2022artificial}. Most studies, however, are considered as data-driven, where the predication is purely based on labeled dataset on limited scenarios. Such models do not perform well with out-of-range data. The problem is exaggerated since signals can be affected by many factors, and incorporating all these elements into a model is quite challenging. With limited datasets, ML-based methods struggle to grasp these non-linear interactions, leading to models that often lack generalizability, robustness, and interpretability.
Recently, there is a trend for applying domain-specific knowledge in ML models that are tailored to the unique requirements of a particular field \cite{karniadakis2021physics}. In the case of radio propagation, ML should be able to  model the \emph{physical process} of the electromagnetic waves propagate through the space, especially interactions with various surfaces and precisely predict MPC or RSS in each TX-RX paths, even in a new environment. In short, it is expected that physics-based ML can provide much faster, efficient, and learnable functions that also yield results to be as close to deterministic simulations as possible.  

\section{Modeling radio propagation with ML: An Overview}

In this section, an overview of ML-assisted channel modeling is provided. To differentiate physics-based model in the next section, we  primarily focus on related works that utilize data-driven approaches. 

\begin{figure}
    \centering
    \includegraphics[width=0.8\linewidth]{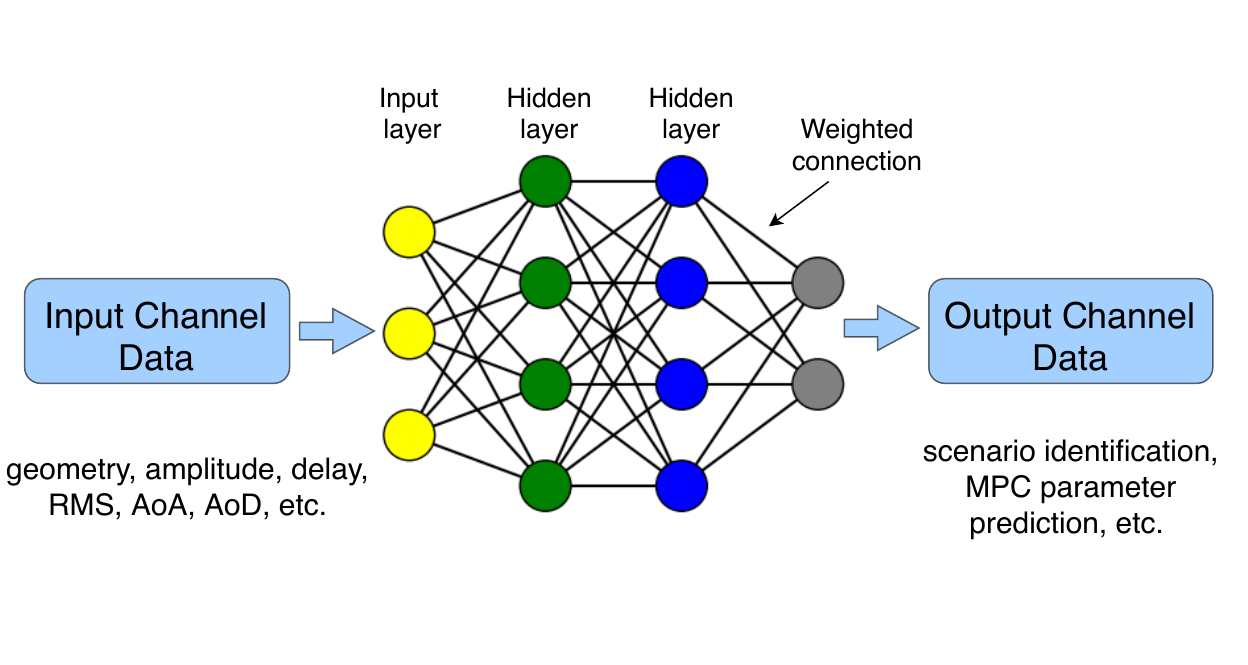}
    \caption{General ML-based radio propagation modeling}
    \label{fig1}
\end{figure}

As shown in Fig. \ref{fig1}, the general architect of ML-assisted channel modeling has three components: input, ML models, and output. The role of ML is to find the mapping between input and output data, which was usually performed by signal processing techniques. In many radio propagation scenarios, the input-output  is non-linear, leading to the difficulty of identifying their explicit relationship. ML-based approach, on the other hand, proved to be highly capable of finding intrinsic patterns and hidden correlations directly from wireless data. The data-driven nature of ML allows for a more accurate and dynamic representation of real-world conditions, accommodating various propagation phenomena like multipath effects, shadowing, and fading.

In literature, the input data  has a diverse format. For example, input (and label) can be signal based parameters such as  amplitude, delay, phase, angle of both AoA and AoD, as well as the geometry data, either detailed 2D/3D layout or stochastic ones like the number of obstacles.  ML, especially deep learning (DL) models are off-the-shelf ones selected from other fields, for example, convolutional neural networks (CNNs) from CV and recurrent neural networks (RNNs) or transformers from natural language processing (NLP). The expected output can be direct prediction of signals parameters, or related classification tasks. In the following, we list some example applications.



\subsection{Radio map estimation with neural networks}
Radio map estimates the propagation pathloss from a TX position to at any point in the planar domain. Also referred as coverage map, it provides important applications in device scheduling and optimal site selection. In a realistic propagation environment where random buildings and street canyons are present, obtaining accurate radio map is challenging. Classical approaches  leverage sparse RSS measurements at certain locations and interpolate other missing points via Kriging kernel or radial basis function (RBF) \cite{romero2022radio}, but suffer from reduced accuracy when handling complicated radio environments. In \cite{levie2021radiounet}, a CNN-based approach is proposed to directly output RSS heatmap when given a TX location and 2D street layout. 
Essentially, this is an image-to-image conversion enabled by a U-Net DL structure. A dataset that contains over 700 street layouts, each with 80 TX positions and the corresponding RSS values from RT are used for training. It is shown that this U-Net neural network (NN) can not only achieve good accuracy but also reveal certain generalizability when given an unseen layouts. Other ML structures have also been explored, \cite{romero2022radio} summarized those works.

\subsection{Scenario identification}
In many applications, it is imperative to identify if the TX-RX communication is LoS or NLoS. For example, if with line-of-sight (LoS) in mmWave systems, beamforming can be designed following the angle of dominant AoA/AoD 
\cite{huang2022artificial}. Historically, such classification problem was formulated as hypothesis testing and solved by likelihood functions. Recently, ML-based method is proved to be more effective in learning key characteristics and distinguishing data in different scenario categories. Specifically, with input of signal amplitude, delay, or angular information,  support vector machine, random forest, and DL such as CNN are common methods to output the LoS or non-LoS (NLoS) identification. A more general identification is to classify propagation scenarios like urban, suburban, highway, etc. This can be helpful for selecting appropriate stochastic channel model and optimal transmission mode. Similarly, with supervised learning, ML methods can map this non-linear mapping via various training signal input. 

\subsection{Application-specific prediction of implicit channel}
For some applications, channel estimation is an intermediate step for performing system optimization. One example is the beamforming for mmWave systems. Indeed, beamformers can be obtained by solving optimization problems with predicted channel, but recent works have utilized ``out-of-the-band'' sensory data from camera and LiDAR to directly learn TX and RX beamformers \cite{xu2022computer}. The underlying mapping is the vision input of environment (implicit channel representation) and the beamformers from optimization problems. This direct mapping is particularly beneficial for large antenna array and dynamic communication scenarios. Another application is to fuse channels of sub-6 GHz signal and few mmWave pilots to predict optimal mmWave beams \cite{gao2021fusionnet}. The use of sub-6 GHz signal is to obtain certain channel spatial information and mmWave pilot can calibrate frequency deviations. 

\subsection{Challenges of ML-based approach}

Data-driven methodologies are inherently model-agnostic, as they depend primarily on the training datasets to optimize ML models through fine-tuning.
However, such approaches have found limited success in channel modeling due to following reasons: 1) While state-of-the-art ML models are capable of capturing complex spatiotemporal relationships, they require a large volume of training data, which is either expensive and labor intensive for measurement campaign, or inconsistent from wireless simulation. 2) Off-the-shelf ML models are not specifically built to tailor the task of channel modeling, hence they often produce unreliable predictions and suffer from performance issues; 3) More importantly, the fundamental challenge is their ability to ``generalize”: to rapidly and accurately solve problems with specifications (geometry, position, TX/RX antenna patterns, and signal frequency) that are beyond those included in the training set. Therefore, their performance largely rely on the quality and quantity of wireless channel data, cannot generalize to out-of-sample scenarios (i.e., those not represented in the training data). 

\section{Physics-informed ML modeling methodologies}


Radio propagation often exhibits a high degree of complexity due to relationships between many physical variables varying across space and time at different scales. Standard ML methodologies shown in the previous section cannot capture such relationships directly from data, especially with sparse measurements. 
A more promising approach is to integrate domain-specific knowledge into ML models such that they can learn and predict generalizable patterns consistent with established governing laws. Towards this direction, physics-informed NN (PINN) has emerged as a foundation tool and been applied in hydromechanics, fluid dynamics, etc \cite{karniadakis2021physics}. 

The goal of PINN-based channel modeling is to create a surrogate NN solver that approximates complicated and computationally intensive simulator such as RT. Different from other surrogate models like Gaussian process or RBF, NN can handle very high dimensional data, making it an ideal choice for our application. In particular, PINN shows the following advantages. 

\subsection{Advantages of PINN in wireless channel modeling}

\subsubsection{Improved generalization and efficiency} 
By embedding signal propagation physical laws into NN, PINNs can generalize better to unseen radio data. 
This is because they are guided by underlying physical principles that govern the system being modeled. Wireless channel exhibits as site-specific and attains to a particular environment with varying objects. Each of these environments introduces unique propagation phenomena that reflected in MPC, resulting in highly-dimensional to highly-dimensional mapping. Properly designed PINN can model and extract such underlying mapping. 

\subsubsection{Channel data efficiency} 
PINNs can be particularly useful in scenarios where data is scarce or expensive to obtain. This is especially true in the high frequency mmWave and terahertz systems. PINN can achieve comparable or superior accuracy with significantly less data by utilizing physical laws as an intrinsic part of their learning process. This reduction in data dependency not only accelerates the training process but also addresses challenges related to data availability and privacy concerns. 


\subsubsection{Interpretability} 
By integrating physical laws, PINNs can offer insights into the underlying mechanisms of the systems they model. This can be very valuable in research and development contexts. The integration of physical laws makes the outputs of PINNs more interpretable. This is highly valuable in wireless channel modeling, as it allows for better understanding of the underlying reasons for model predictions, facilitating decision-making and system optimization.

\subsection{Recent representative works}
In the past year, several works have explored to generalize channel modeling with PINN. In the following, we briefly summarize these works, particularly from the modeling aspect. 

\subsubsection{Generalizable radio map prediction} 
A more generalized radio map prediction is presented in \cite{9771088}, \cite{seretis2022toward}. For a given 2D indoor layout that has pixels representing concrete, wood, and glass using different RGB color, their proposed \emph{EM DeepRay} can accurately replicate the pathloss results of a RT. The key idea is to use CNN, and each input feature (permittivity, conductivity, distance, and free-space law) are encoded with convolutional filters. The overall network structure is similar to U-Net. Even with complex and unseen layout, \emph{EM DeepRay} can produce heatmap with small mean absolute error (MAE). However, 2D layout cannot reflect real-world scenarios, and their system only considers empty rooms with limited materials shown in pixel values. 

\subsubsection{Environment-aware spatial signal prediction}
In CV, neural radiance fields (NeRF) is used to trace optical rays. \emph{NeRF} regards each pixel of an image as a result of one ray tracing. It captures a few images in the scene from different angles as input then trains an multi-layer perceptron (MLP) to fit the scene-dependent optical radiance field.  \cite{10.1145/3570361.3592527} utilizes the idea of implicit representation from \emph{NeRF} and extend to radio rays. The main difference is that radio rays may bounce more than once, therefore, multi-reflection can be regarded as a new TX that ``retransmit'' a combined signal received from all possible paths. The proposed $\emph{\text{NeRF}}^2$ can output the spatial spectrum for any TX-RX pair in the trained scene. Evaluation results show that $\emph{\text{NeRF}}^2$ can predict accurate multipath profile at any RX location. Although trained with PINN, the drawback is that all results are constrained in the trained scene, same as \emph{NeRF}. It needs re-training when given new 3D scene, which limits the generalizability. 

\subsubsection{Ray-surface interaction with NN}
As mentioned before, RT may suffer from computational issues, mainly in the realization of ray's interaction with objects. 
\emph{WiNeRT} is proposed as a neural wireless simulator, which serves as the ML surrogate model to replace signal's interaction with surface of each encountering object \cite{orekondy2022winert}. Same as RT, the TX omni-directionally launches 100,000 rays and \emph{WiNeRT} learns the mapping of an incident ray with some direction and power to an updated outgoing attenuated ray, and accurately reflect surface materials and spatial coordinates by treating objects as a set of vertex. The system will collect all rays that intersect with the RX coordinate. \emph{WiNeRT} is  much faster than RT, and at the same time, can implicitly learn material EM property, free-space law (based on ray travel distance), and more general geometric interactions. However, it needs to shoot a fix number of rays, even with given 3D layouts. Besides, \emph{WiNeRT} focuses more on reflection, lacking the modeling of scattering and diffraction. 

\subsubsection{mmWave beam prediction with segmentation} To enable high data rate in vehicular scenarios, \cite{10.1145/3570361.3613291} proposed \emph{mmSV} that uses the street view pictures as input, then reconstructs the 3D environment with the material assignment by semantic segmentation to assist vehicles in finding mmWave reflections from the environment in real-time. \emph{mmSV} consists of two major components: building surface material classification and environment-driven RT. The former automatically identify EM properties of common outdoor object surfaces and the latter facilitates the search for optimal beam directions towards RX. Experimental results show a high signal-to-noise ratio (SNR) can be achieved with this PINN workflow, even under vehicle dynamics. Due to dataset and computation complexity, this work may not work well for fine-grained signal predictions, such as MPCs. 
Table. \ref{tab:my_label} summarized related works that introduced in this paper.

\begin{table}[h]
    \centering
    \begin{tabular}{| p{20mm} | p{12mm} | p{30mm} | p{50mm} | p{20mm} |}
    \hline
       \textbf{Related Works} & \textbf{PINN?} & \textbf{Network Structure} & \textbf{Input-Output Data Type} & \textbf{Generalizability}  \\
        \hline
        \cite{levie2021radiounet} & No &  U-Net (CNN) &  \emph{In}: \{2D layout, TX position\}, \emph{Out}: \{RSS heatmap\}  & Low  \\ 
        \hline
        \cite{xu2022computer} & No &  CNN-LSTM & \emph{In}: \{images, signal features\}, \emph{Out}: \{beamformers\} & Low \\
        \hline
        \cite{gao2021fusionnet} & No &  CNN & \emph{In}: \{sub-6 GHz signal features, few mmWave pilot\}, \emph{Out}: \{beamformers\} & Low \\
        \hline
        \cite{9771088}, \cite{seretis2022toward}  & Yes & CNN & \emph{In}: \{indoor geometry, TX position\}, \emph{Out}: \{RSS heatmap\}  & Medium \\
        \hline
        \cite{orekondy2022winert} & Yes & MLP &   \emph{In}: \{3D layout, TX position\}, \emph{Out}: \{MPCs\} & High \\
         \hline
         \cite{10.1145/3570361.3613291} & Yes & GoogLeNet, DeeplabV3  & \emph{In}: \{street view images, TX position\}, \emph{Out}: \{beamformer (AoD)\}  & High \\
        \hline
         \cite{10.1145/3570361.3592527} & Yes  & MLP & \emph{In}: \{indoor point cloud, TX position\}, \emph{Out}: \{AoD\}  & Medium \\
         \hline
    \end{tabular}
    \caption{Summary of related works}
    \label{tab:my_label}
\end{table}

\subsection{PINN design methodologies}

\subsubsection{Proposed architect}

\begin{figure}
    \centering
    \includegraphics[width=0.9\linewidth]{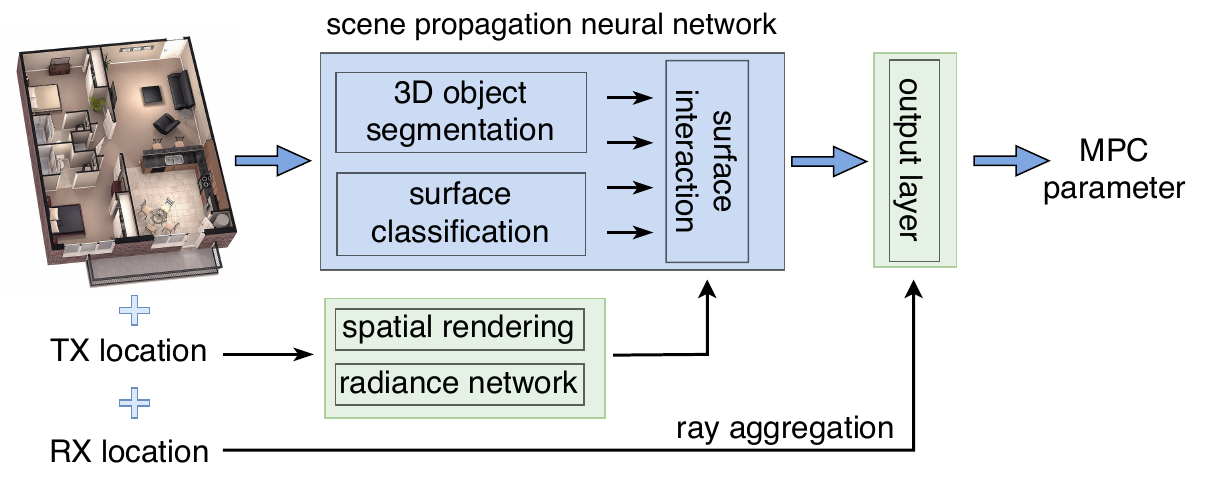}
    \caption{Proposed PINN-based channel modeling architect}
    \label{fig:pinnl}
\end{figure}

Intuitively, wireless channel model is site-specific, therefore, detailed 3D indoor layout that contains furniture, appliances, and building objects should be used as input. As shown in Fig. \ref{fig:pinnl}, TX and RX locations are separately encoded. Indoor 3D layout can be obtained from various sources, for example the depth camera, LiDAR point cloud, or their fusion.  A scene propagation NN is utilized to process 3D input and obtain semantic understanding of the environment. Based on the TX location, a spatial and radiance NN is used to guide ray propagation through the space. At the RX side, ray is aggregated at RX location. The processed intermediate results are then sent to the output layer for MPC parameter prediction. Through MPC, LoS/NLoS classification and RSS heatmap can also be derived.    

\subsubsection{Scene propagation NN}
This network consists of three components. The first is 3D object segmentation for item recognition, particularly major reflectors and edges. It is also able to extract dimension information. Many 3D semantic segmentation algorithms can be applied from CV. 2) Surface classification that can recognize covering material of each segmented object, for example, the refrigerator of metal, the ceiling of vinyl or dry wall, table of wood, etc. 3) The surface interaction that receives inputs from processed semantic understanding and spatial rending via the path out of TX location. This part is a surrogate model similar to \cite{orekondy2022winert}.   
\subsubsection{NN path rendering}
To simulate ray marching of RT, TX location is used as the input to the path rendering NN that makes local evaluations of each ray. Specifically, as shown in \cite{orekondy2022winert,10.1145/3570361.3592527}, rays can bounce several times in the space, an efficient tracking approach is to map each incident ray with direction and power to a new outgoing ray, by treating the point-of-contact as a new ``TX''. Thus, the bounce of rays becomes an iterative process, reducing the complexity of global knowledge. Through surface interaction, the implicit non-linear mapping of surface material and spatial locations will be rendered to the output layer.  

\subsubsection{PINN loss function}
Loss function is an important design metric to help  ML models capture generalizability. The loss function should contain at least two parts \cite{10.1145/3514228}: $MSE_u = \text{Loss} (Y_{true}, Y_{pred}) + \lambda R(\mathbf{w})$  and $MSE_f = \gamma \text{Loss}_{PHY}(Y_{pred})$. The former is a commonly used loss function in ML, representing supervised error between predicted value and true label such as the MAE of true and predicted AoAs, and a regularization term like norm function to limit the model complexity. The latter is to ensure consistency with physical laws and it is weighted by a hyperparameter $\gamma$. $ \text{Loss}_{PHY}(Y_{pred})$ is application-specific, in this case can be a physics-based penalty that ensures the power of rays attenuates with the distance traveled or reflected angle from incident ray. 
Physics-informed loss function has several benefits, mostly for reducing search space and accommodating unlabeled data. 

\section{A case study with physics-informed indoor propagation modelling}
In this section, we provide a case-study for precise indoor MPC prediction with PINN. This work has a complete workflow from synthetic data generation, PINN algorithms, to initial results that show the effectiveness of PINN modeling. 

\subsection{3D modeling and segmentation dataset}
\begin{figure}
    \centering
    \includegraphics[width=0.9\linewidth]{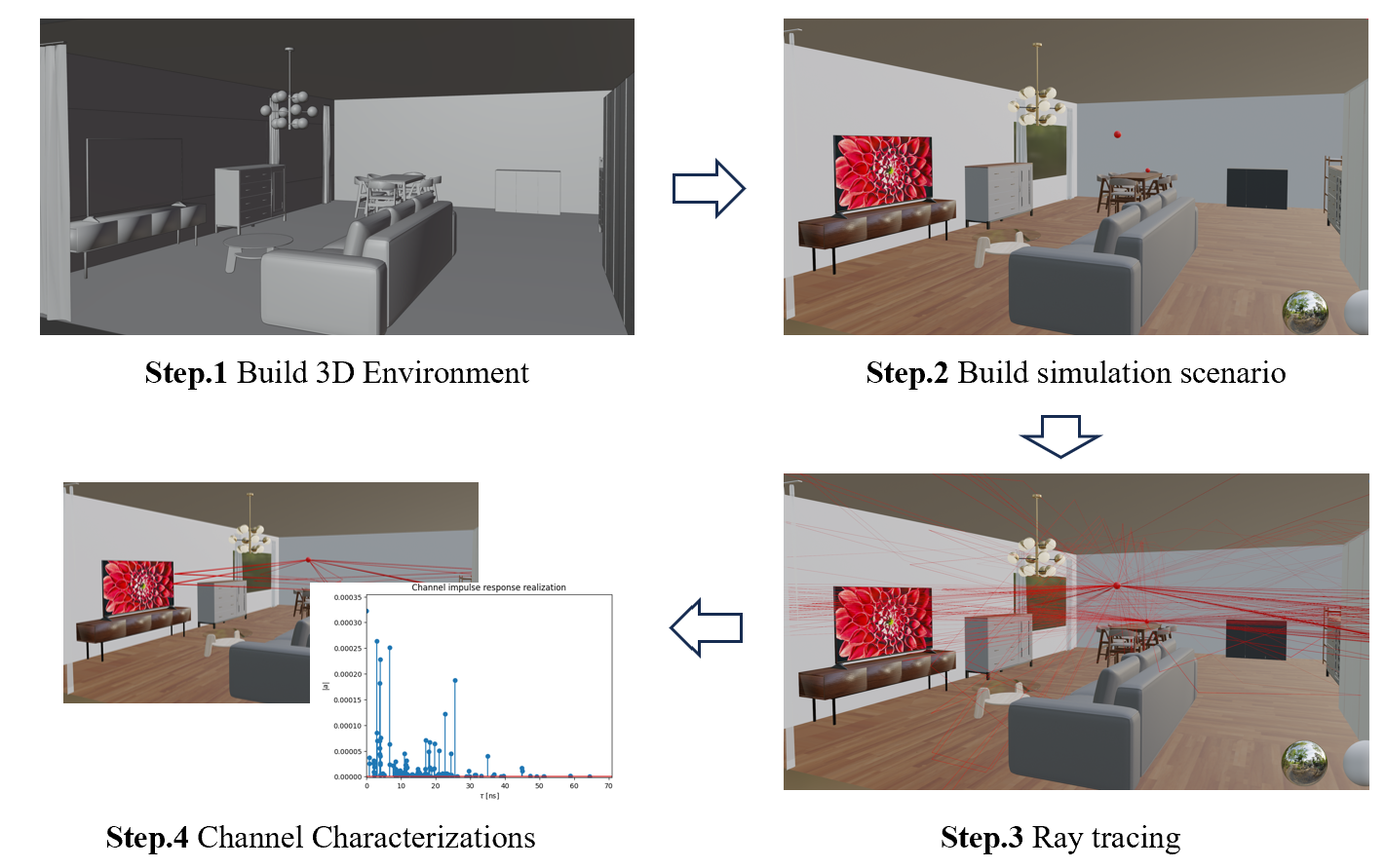}
    \caption{Synthetic indoor radio dataset generation workflow}
    \label{fig:dataset}
\end{figure}

There exist wireless dataset from measurements or simulation that aim to capture the environment and signal relationship, but they 
often neglect detailed elements of indoor environments like potential reflective surfaces, scattering zones, and diffraction edges. Consequently, models developed for simplistic 2D or 3D environments and trained on basic data may not capture these complex features. 
Thus, we built \emph{WiSegRT} \cite{wiseg}, which aims to provide a fine-grained environment and co-located channel parameters, this means researchers can not only obtain comprehensive radio propagation data from the elaborately created environment, but also generate corresponding visual data for vision-aided radio systems. Fig. \ref{fig:dataset} shows the synthetic data generation workflow. The  GPU-accelerated and differentiable \emph{Sionna} RT  is used. The process starts from indoor 3D scene modeling with \emph{Blender}, each object is covered with high resolution mesh and segmented with label. Objects are assigned with materials of different conductivity and permittivity. The scene and locations of TX and RX are loaded into \emph{Sionna}, which launches 100,000 rays  omnidirectionally. Rays are limited to bounce for less than 4 times in the space to reduce complexity. The RX will collect ``valid'' rays that arrives at its antenna. Finally, attenuation and delay of each valid ray are calculated to obtain the CIR profile. We have created over 10 realistic indoor scenes, each with over 30 TX and 3,000 uniform RX positions.  

\subsection{Result 1: Pathloss variation comparison}

\begin{figure}
    \centering
    \includegraphics[width=0.6\linewidth]{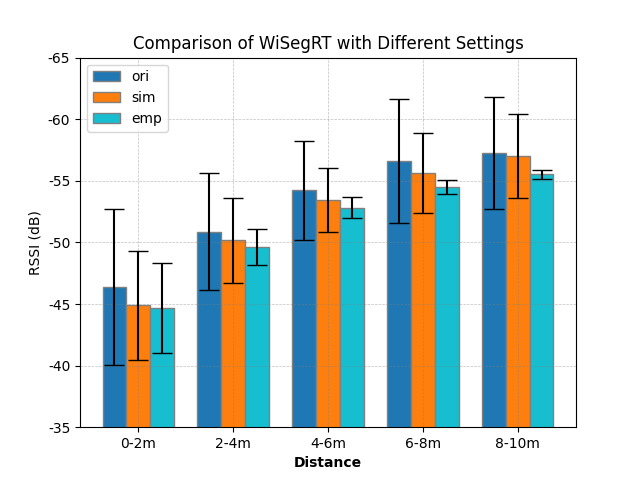}
    \caption{WiSegRT shows more system dynamic compared with existing  datasets  \cite{wiseg} }
    \label{fig:result1}
\end{figure}

Figure \ref{fig:result1} illustrates a comparative analysis of the mean RSS across various TX-RX distance brackets. The RSS for each TX-RX pair is derived by aggregating the total power received from all MPCs. While the mean RSS values across the three evaluated scenarios exhibit marginal variations when measured on a decibel scale, a marked distinction is observed in their standard deviation (SD). This SD is largely caused by effects of environmental multipath phenomena and obstructions in the LoS when categorizing TX-RX pairs based on distance. The SD is a critical metric reflecting the fidelity with which each scenario replicates real-world radio propagation characteristics. ``original" (ori), ``simulated" (sim), and
``empirical" (emp) stands for dataset with high definition 3D objects and dedicated covering materials, coarse objects and no covering material, and only layout (no objects or materials), respectively. 
A reduction of SD relative in ``sim'' scenario shows impacts due to the absence of detailed 3D object modeling and electromagnetic properties of materials on RSS distribution. By removing all interior objects, the simulated RSS values in the ``emp" scenario exhibit a substantially reduced SD at distances beyond two meters, indicating that such an ``empirical" configuration does not accurately represent the RSS distribution encountered in real-world environments.

\subsection{MPC prediction with segmentation and knowledge distillation}
\begin{figure}
    \centering
    \includegraphics[width=0.8\linewidth]{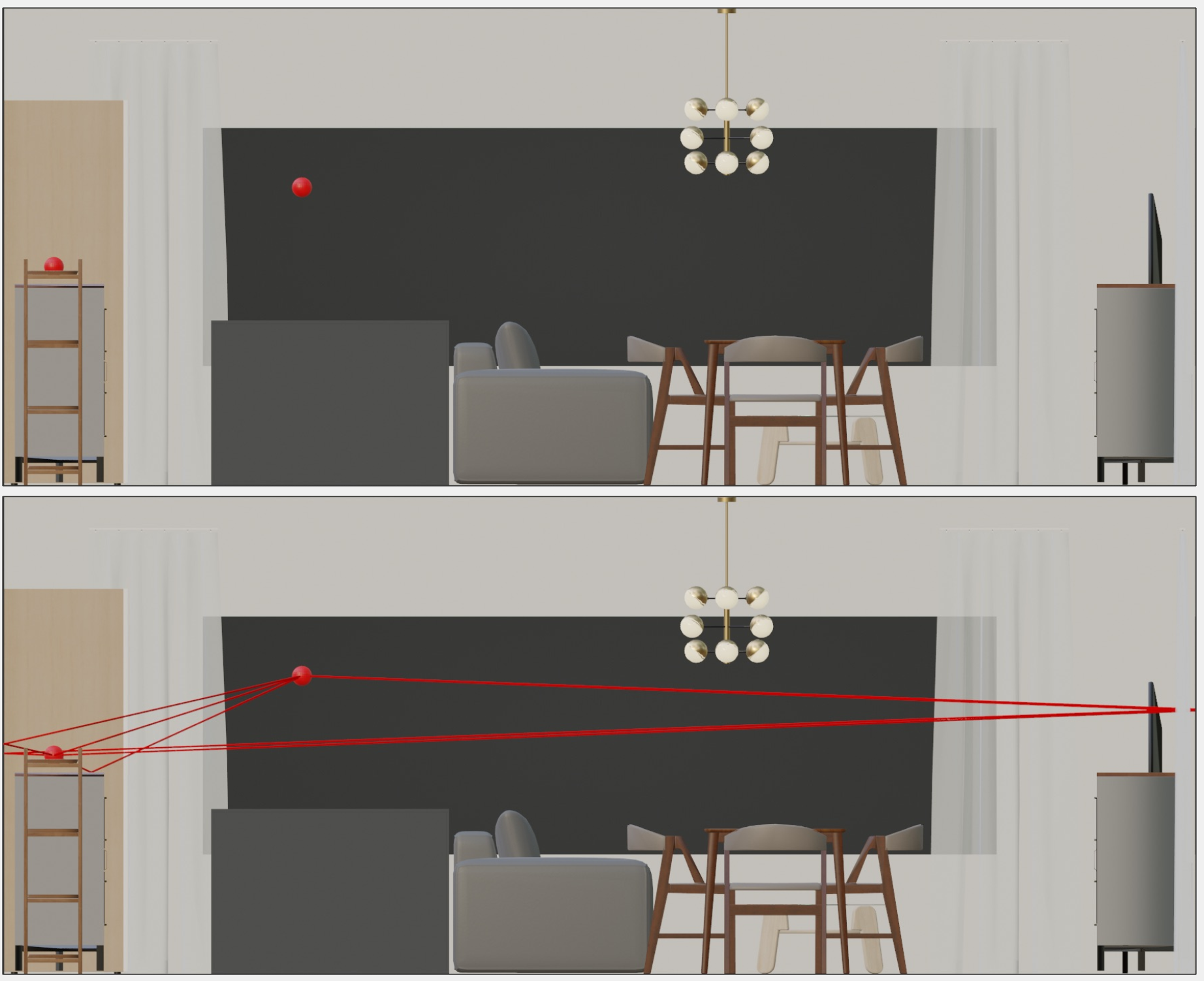}
    \caption{MPC prediction with segmentation and knowledge distillation}
    \label{fig:result2}
\end{figure}

We have conducted initial investigation on the direct prediction of MPCs. Generative adversarial network (GAN) is applied where a generator is used to generate predicted MPCs based on the input of an indoor snapshot and TX-RX position, a discriminator makes binary decision on whether the MPC is real or fake based on segmentation information and propagation knowledge distillation.  In the context of GANs, achieving equilibrium refers to a balanced state where both the generator and the discriminator mutually benefit and improve from their game. However, the use of a pre-trained then fine-tune GAN proved not effective in our first run, as it is unable to learn the intrinsic relationship of ray's distribution and objects in the space, which results discriminator to outperform generator. An improved Eq-GAN-SA \cite{wang2022improving} is applied. It uses an PINN-based attention map from discriminator to guide the process of generator with manual tuning of each input feature. The result is shown in Fig. \ref{fig:result2} where we cap the maximum ray interaction at 2.

\section{Challenges and future directions}

\subsection{Challenges for PINN channel modeling}

\subsubsection{Limited high resolution dataset} The first challenge is the lack of measurement data, not to mention the high quality and resolution ones. Collecting data is time consuming and labor intensive, and for high frequency signals, the devices are expensive and hard to setup. Environment-aware channel data are multi-modality from sources like vision (camera or LiDAR), as well as the wireless domain. For indoor scenarios, MPCs should have a temporal resolution of 2 ns or better. Besides, environmental diversity that includes more scene setups and layouts are ideal for PINN training. 

\subsubsection{Gaps between simulation and measurement} In today's wireless channel research, due to scarcity of measurement data, simulator is usually applied to generate both training and testing data. However, there exists performance gaps between them due to following reasons: 1) Non-linear wireless RF components. For example, the power amplifier may cause signals to develop non-uniform distribution in the space. 2) RT simulators cannot take all propagation characteristics into consideration. In the case of \emph{Sionna}, it fails to model refraction, i.e., the signal penetration through objects, and has limited ability to model scattering. For high frequency signals, scattering caused by non-smooth surface can have substantial impacts.  3) Practical receiver has limited temporal, frequency, and spatial resolution. 

\subsubsection{Deep integration of PINN with radio propagation} Numerous factors can influence wireless channel, resulting in a task that is extremely dynamic and heavily dependent on the surrounding environment. However, incorporating all propagation factors into PINN architectures leads to increased model complexity and high computational overhead, which diminish its advantages. Besides, a good model should balance data-driven and physics, as also can be seen from Fig. \ref{fig:pinnl}, scene propagation NN relies on data-driven models for intermediate features, while radiation network is based on PINN. 

\subsection{Future opportunities}
The recent studies presented in Section III.B are exploratory efforts in channel modeling using PINN. Progress in ML models, coupled with the increasing availability of datasets, is expected to significantly advance this area of research.
\subsubsection{Digital twin dataset generator} Digital twin (DT) is a virtual model designed to accurately reflect a physical system. For wireless research, it is increasing popular to design a DT that can produce data indistinguishable of a physical system \cite{khan2022digital}. The challenge of scarcity data can be partially addressed by DT generators, which can ensure a flexible, cost-effective, and consistent workflow. Besides, DT can simulate complex environment and obtain corresponding labeled data to help the training of PINN. 

\subsubsection{Large ML models} Recent progress on large ML models, such as NLP and CV  models, prove to be very effective for solving complex tasks, such as reasoning and multi-modality data processing. While PINN-based channel modeling approach can reduce both model and data size, another possible direction is to develop PINN-based large ML models, especially when the input data is from several sources (RGB camera, point cloud, etc). As shown in Fig. \ref{fig:pinnl}, the scene propagation NN can be implemented with larger ML models for fine-grained scene understanding and feature extraction, while the rendering network can be based on PINN concept. 

\section{Conclusion}
Channel modeling plays a crucial role in the development of wireless systems hence has attracted  significant research attention. Recent years have witnessed the increasing utilization of data-driven methods to simplify modeling process and provide accurate channel prediction. In this work, we first briefly summarized data-driven methods and listed their drawbacks. We then presented the concept of PINN-based modeling and summarized representative works. Through one case study, we have shown that PINN for channel modeling has the potential to be generalizable, interpretable, and robust. A general architect of PINN methodology is provided that aims to guide future model design. This work is concluded with challenges and research directions. 

\bibliographystyle{IEEEtran}
\bibliography{sample}

\end{document}